\documentclass[twocolumn,prx,preprintnumbers,showpacs,amsmath,amssymb,superscriptaddress]{revtex4-1}
\usepackage[pdftex, colorlinks, citecolor=blue]{hyperref}   
\usepackage{graphicx}
\usepackage{dcolumn}
\usepackage{threeparttable}
\usepackage{multirow}
\usepackage{txfonts}
\usepackage{xcolor}
\usepackage{amssymb}
\usepackage{amsmath}
\usepackage{latexsym}
\usepackage{epsfig}
\usepackage{amsbsy}
\usepackage{array}
\usepackage{tabularx}

\newcommand{\ket}[1]{\mid #1 \rangle}

\newcommand{\sais}{\ifmmode\mbox{\c{S}}\else\c{S}\fi{}a\ifmmode\mbox{\c{s}}\else\c{s}\fi{}\ifmmode
\imath \else \i \fi{}o\ifmmode \breve{g}\else \u{g}\fi{}lu~}

\begin{document}

\title{Doping-induced insulator-metal transition in the Lifshitz magnetic insulator NaOsO$_3$}

\author{Sabine Dobrovits}
\affiliation{University of Vienna, Faculty of Physics and Center for
Computational Materials Science, Sensengasse 8, A-1090 Vienna, Austria}

\author{Bongjae Kim}
\affiliation{University of Vienna, Faculty of Physics, Sensengasse 8, A-1090 Vienna, Austria}
\affiliation{Department of Physics, Kunsan National University, Gunsan 54150}

\author{Michele Reticcioli}
\affiliation{University of Vienna, Faculty of Physics and Center for
Computational Materials Science, Sensengasse 8, A-1090 Vienna, Austria}

\author{Alessandro Toschi}
\affiliation{Institut f\"{u}r Festk\"{o}rperphysik, Technische Universit\"{a}t Wien, Vienna, Austria}

\author{Sergii Khmelevskyi}
\affiliation{Institute for Applied Physics and Center for Computational Materials Science, Technische Universit\"{a}t Wien, Vienna, Austria}

\author{Cesare Franchini}
\email{cesare.franchini@univie.ac.at}
\affiliation{University of Vienna, Faculty of Physics and Center for
Computational Materials Science, Sensengasse 8, A-1090 Vienna, Austria}
\affiliation{Dipartimento di Fisica e Astronomia, Universit\`{a} di Bologna, 40127 Bologna, Italy}

\date{\today}

\begin{abstract}

By means of first principles schemes based on magnetically constrained density functional theory and on the band unfolding technique we study the effect of doping on the conducting behaviour of the Lifshitz magnetic insulator NaOsO$_3$. Electron doping is treated realistically within a supercell approach by replacing sodium with magnesium at different concentrations. Our data indicate that by increasing carrier concentration the system is subjected to two types of transition: (i) insulator to bad metal at low doping and low temperature and (ii) bad metal to metal at high doping and/or high-temperature.
The predicted doping-induced insulator  to metal transition (MIT) has similar traits with the temperature driven MIT reported in the undoped compound. Both develops in an itinerant background and exhibit a coupled electronic and magnetic behaviour characterized by the gradual quanching of the (pseudo)-gap associated with an reduction of the local spin moment. Unlike the temperature-driven MIT, chemical doping induces substantial modifications of the band structure and the MIT cannot be fully described as a Lifshitz process.

\end{abstract}

\maketitle

\section{Introduction}

Metal-insulator transitions (MIT) have long been a focal point of condensed matter physics~\cite{Imada1998} due to the inherent conceptual complexity, which has stimulated the development of many theories~\cite{Mott1968}, and to the possibility to control the (reversible) suppression of electrical conductivity in technological applications~\cite{Yang2011}. Understanding and describing MIT is a considerable task. In the most simple scenario, metals and insulators can be distinguished within the non-interacting Wilson's picture based on the filling of the electronic bands~\cite{Wilson458,Wilson277}. Wilson's approach correctly predicts the insulating nature of fully-filled/empty $d$-bands transition metal oxides (TMO) such as SrTiO$_3$ ($d^0$) and,  to some extent, Cu$_2$O ($d^{10}$) and LaCoO$_3$ ($t_{2g}^6$), but breaks down for partially filled $d$-bands TMOs like NiO and many others~\cite{Boer}. With his pillar works,  Mott has resolved this limitation by considering the effect of electron-electron correlation and formulated one of the most influential paradigm in solid state physics, the Mott insulator~\cite{Mott1949,Mott1968}, that is still the subject of intense research nowadays.

The recent discovery of novel types of MITs in spin-orbit coupled 5$d$ TMOs, such as the Dirac-Mott regime in Sr$_2$IrO$_2$~\cite{Kim2008,Kim2009,Jackeli2009,Liu2015,Liu2018} and the magnetically itinerant phases of 5d$^3$ osmathes NaOsO$_3$~\cite{Calder2012,Jung2013,Middey2014,Kim2016} and Cd$_2$Os$_2$O$_7$~\cite{Mandrus2001,Yamaura2012,Hiroi2015} has given additional momentum 
to the research on correlated materials.
The experimental data evidencing the MIT in NaOsO$_3$ and  Cd$_2$Os$_2$O$_7$ are difficult to decipher and rationalize. Both compounds show a continuous, second-order temperature-driven transition accompanied by the onset of a magnetic order~\cite{Calder2012,Mandrus2001}.
Initially, the  BCS like-gap inferred by infrared spectroscopy studies~\cite{Padilla2002,Vecchio2013} and the observation that the N\'{e}el temperature coincides with the critical MIT temperature suggested a Slater-type mechanism. In fact, in a Slater insulator the onset of the insulating regime is combined with the simultaneous formation of a long-range antiferromagnetic (AFM) order~\cite{Slater,Calder2012,Mandrus2001}. 
However, it was soon realized that a purely Slater scenario is incompatible as a full explanation of the experimental observations in NaOsO$_3$, in particular of their evolution as a function of $T$.  In fact, the high-degree of magnetic fluctuations and electron itinerancy observed in these Osmathes are better captured by an alternative Lifshitz-like picture involving a rigid upward (downward) shifts of electron (hole) bands~\cite{Lifshitz, Hiroi2015, Kim2016}.

The Lifshitz MIT in NaOsO$_3$ is driven by temperature. 
At high temperatures NaOsO$_3$ is a paramagnetic metal, with strongly fluctuating magnetic moments. By decreasing temperature the magnetic fluctuations are gradually frozen, leading to the continuous vanishing of holes and electrons pockets in the Fermi surface, that do not involve any substantial modification of the underlying band topology \cite{Kim2016}.
More precisely, at the N\'{e}el temperature ($T_N$ $\approx$ 410~K) a pseudogap develops, which can be related to the attenuation of rotational spin fluctuations and the formation of a long-range AFM ordering, and the system enters a bad metal regime~\cite{Shi2009} characterized by longitudinal modulations of the spin moment~\cite{Kim2016}. Further lowering of the temperature favors a the full opening of an insulating gap at about $T=30$~K, corresponding to the eventual freezing of the longitudinal fluctuations.

There are two additional peculiar aspects of the MIT in NaOsO$_3$. First, despite being in a nominally $t_{2g}^3$ configuration the ordered moment is only 1~$\mu_{B}$~\cite{Calder2012} due to an high degree of $p-d$ hybridization which place the system close to an (electronic and magnetic) itinerant limit~\cite{Jung2013,Kim2016,Calder2017}; in additional even though the orbital moment is formally quenched ($L_{{\rm{eff}}}=0$, nominally 5$d^3$ configuration~\cite{good}), spin-orbit coupling effects are surprisingly important as they cause a renormalization (weakening) of the electron-electron correlation~\cite{Kim2016} and a large magnetic anisotropy energy~\cite{Singh2018}.

\begin{figure}[h!]
\includegraphics[width=0.45\textwidth]{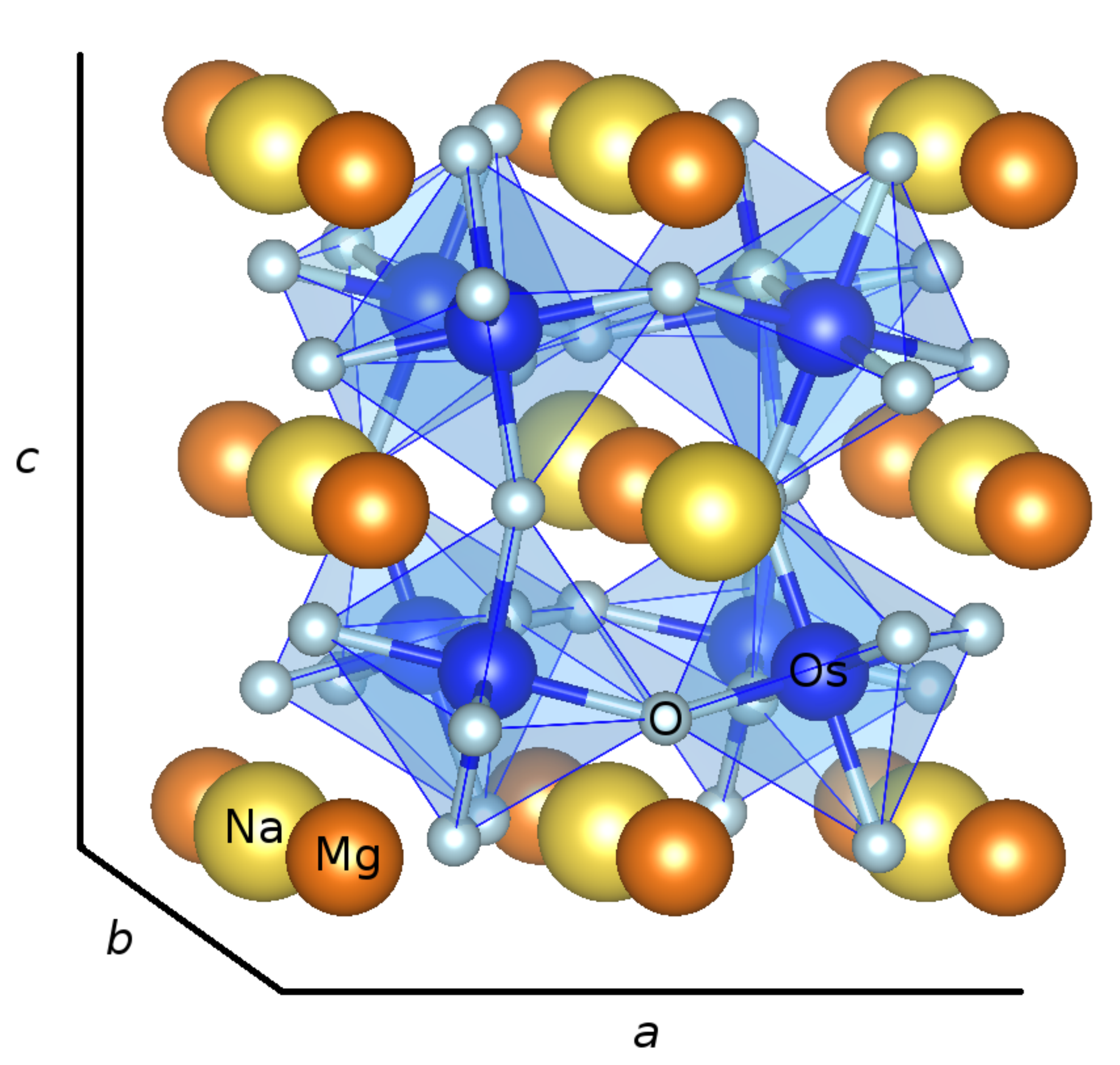}
\caption{(Color online) Schematic view of the ${2}\times{2}\times{2}$ supercell adopted in this study to model Na$_{1-x}$Mg$_x$OsO$_3$. The polyhedron indicate the OsO$_6$ octahedra and the remaining large and small spheres represent Na and Mg atoms, respectively. This sketch shows the half-doping regime ($x=0.5$) characterized by a uniform chessboard distribution of Na/Mg ions with chemical formula Na$_{4}$Mg$_4$Os$_8$O$_{24}$. For the other doping concentrations different type of inequivalent Mg distribution have been tested.}
\label{fig:structre}
\end{figure}

In general, the ground state of a system can be perturbed by different means including temperature~\cite{Imada1998}, doping~\cite{Liu2016,Franchini2005}, pressure~\cite{Sanna2004,He2012}, strain/heterostructing~\cite{Kim2017}, and dimensionality~\cite{Liu2018} to name the most effective stimuli. 
In this study we inspect the possibility to control the MIT in NaOsO$_3$ via chemical doping by means of first principles calculations.
Doping effects can be modeled by following a variety of routes which could involve: (i) a rigid shift of the band (rigid doping); 
(ii) a controlled change of the number of valence electrons (preserving charge neutrality via an homogeneous background charge); (iii) the virtual crystal approximation or (iv) realistic doping via chemical substitution. We follow this latter strategy by replacing Na with Mg at different concentration within a supercell approach (a sketch of the adopted Na$_{1-x}$Mg$_x$OsO$_3$ supercell is shown in Fig.~\ref{fig:structre}).
The results indicate that by injecting a progressively larger amount of excess electrons the systems undergoes a MIT associated with an almost linear decrease of the ordered moment, volume reduction, and characterized by a transition between a bad-metal regime (pseudogap between the partially occupied bottom of the conduction band and the fully filled $t_{2g}$ valence manifold) to a full metal state where the magnetic gap is fully quenched.

The manuscript is organized as follows.
We start from a brief description of the technical setup. Subsequently, we present and discuss the results on the doping-induced MIT in NaOsO$_3$ and draw a general phase diagram showing the intersection between the pseudogap and metallic regime as a function of doping, size of the magnetic moment and temperature. 

\section{Computational details}

Our first-principles calculations were performed using the
projector augmented wave method (PAW)~\cite{PhysRevB.50.17953} as implemented in the Vienna
\emph{Ab initio} Simulation Package (VASP)~\cite{PhysRevB.47.558, PhysRevB.54.11169}.
The plane-wave cutoff for the orbitals was set to 400 eV and to sample the Brillouin zone a
3$\times$3$\times$3 $k$-point grid was used, generated according to Monkhorst-Pack scheme.

All calculations were performed using a fully relativistic setup with the inclusion
of SOC in the framework of the DFT+U~\cite{PhysRevB.57.1505, Dudarev2018} with an effective $U_{eff} = U-J = 0.6$~eV~\cite{Kim2016} and using the PBE parametrization of the exchange-correlation functional.

The unit cell of NaOsO$_3$ contains four formula units consisting of 20 atoms. With respect to the ideal cubic perovskite  (1$\times$1$\times$1) unit cell the magnetic unit cell of undoped NaOsO$_3$ is constructed by a 45$^{\circ}$ rotation around the $y$ axis and a doubling of the $b$ lattice parameter, i.e. ($\sqrt{2}\times{2}\times\sqrt{2}$), with experimental lattice parameters $a$=5.3842~\AA, $b$=7.5804~\AA, $c$=5.3282~\AA~\cite{Shi2009}. The resulting orthorhombic $Pnma$ structure is subjected to small internal geometrical distortions which leads to slightly different Na-O distances.

Realistic electron doping was modeled by chemical substitution of Na with Mg in {2}$\times$2$\times${2} 
supercells containing 8 Na$_{1-x}$Mg$_x$OsO$_3$ formula unit (40 atoms) for different Mg concentrations $x$=0.125, 0.25, 0.375, 0.5  and considering different configurations of the dopants. All supercells were fully relaxed including both volume (lattice parameters) and internal atomic positions.

To analyze the effects of doping on the energy band structure, we projected the states of the supercell onto the states of the primitive cell by adopting the unfolding method~\cite{Boykin,Popescu2010,Popescu2012} recently implemented in VASP~\cite{Eckhardt2014,Reticcioli2015}.
The projection $P_{\vec{K}m}(\vec{k})$, also known as Bloch character, is calculated as
\begin{equation}
P_{\vec{K}m}(\vec{k}) = \sum_n|\langle\Psi_{\vec{K}m}|\psi_{\vec{k}n}\rangle|^2~.
\label{eq:bloch}
 \end{equation}
where $\ket{\Psi_{\vec{K}m}}$ and $\ket{\psi_{\vec{k}n}}$  are the eigenstates of the supercell and primitive cell, respectively, $K$ and $k$ the respective wave vectors
and $m$ and $n$ are energy band indexes (more details on the method can be found in Ref.~\onlinecite{Liu2016, Reticcioli2017}).

\section{Results and discussions}

\begin{figure}[h]
\includegraphics[width=0.49\textwidth]{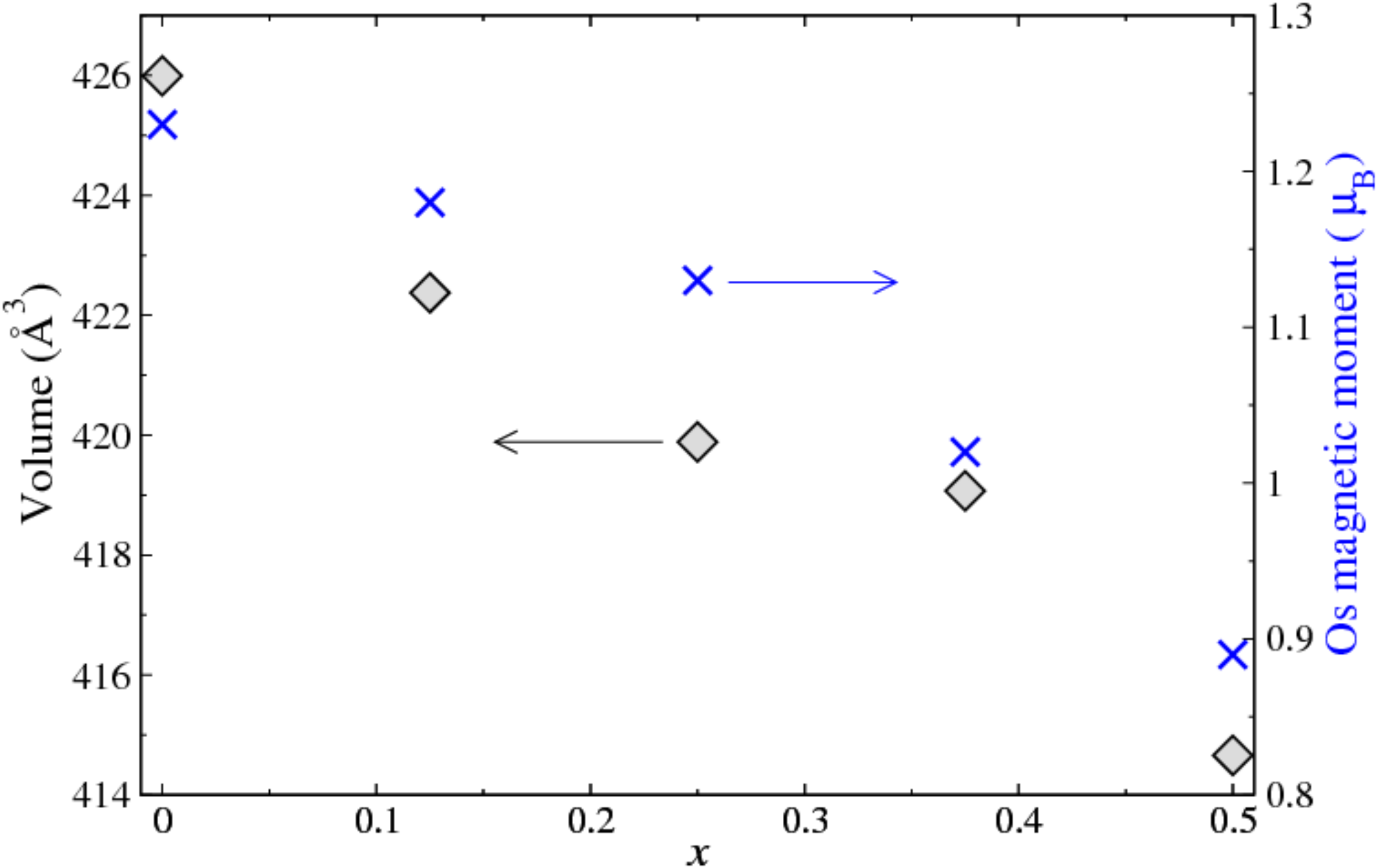}
\caption{Variation of the volume  [left scale] and of the ordered Os spin moment [right scale] as a function of Mg concentration $x$. 
Data averaged over all studied configurations.}
\label{fig:volume}
\end{figure}

As a starting point we inspect the evolution of the structural and magnetic properties upon doping. In fact, considering the strong spin-phonon interaction in NaOsO$_3$~\cite{Calder2015}, it is expected that the effect of doping should not be limited to purely electronic effects, but rather involve a concerted change of volume and local moment. As a consequence of the smaller atomic radius, Mg-substitution causes a gradual decrease of the volume, as shown in Fig.~\ref{fig:volume} and tabulated in Tab.~\ref{Tab:data}. Going from the undoped ($x=0$) to the half-doped ($x=0.5$) sample the volume is squeezed by about 3\%. This structural change is associated with a huge ($\approx 30\%$) lowering of the local spin moment from 1.23~$\mu_B$ ($x=0$) to 0.89~$\mu_B$ ($x=0.5$). In  NaOsO$_3$, this should reflect an increase degree of electron and magnetic itineracy. Thus, already at this stage, we expect to have a coupled magnetic-electronic transition induced by electron doping, in analogy with the T-driven MIT, where the collapse of the gap is associated with a gradual quenching of the local spin moment (see introduction).

\begin{table}[t]
\footnotesize
\caption {Collection of structural properties [average volume ($\AA^3$~per supercell) and lattice constants ($\AA$)] and average Os magnetic moment, in $mu_B$) as a function of Mg doping $x$.
The structural data refer to the {2}$\times$2$\times${2} supercell adopted in the calculations.
}
\begin{ruledtabular}
\begin{tabular}{lccccc}
  $x$    & V & a & b & c & m \\
  \hline
   0     & 426.00  & 7.530  &  7.513  &  7.530 & 1.23 \\
   0.125 & 422.37  & 7.502  &  7.502  &  7.506 & 1.18  \\
   0.25  & 419.89  & 7.496  &  7.493  &  7.477 & 1.13 \\
   0.375 & 419.07  & 7.502  &  7.514  &  7.436 & 1.02 \\
   0.5   & 414.66  & 7.469  &  7.504  &  7.400 & 0.89   
\end{tabular}
\end{ruledtabular}
\label{Tab:data}
\end{table}

\begin{figure}[h!]
\includegraphics[width=0.315\textwidth]{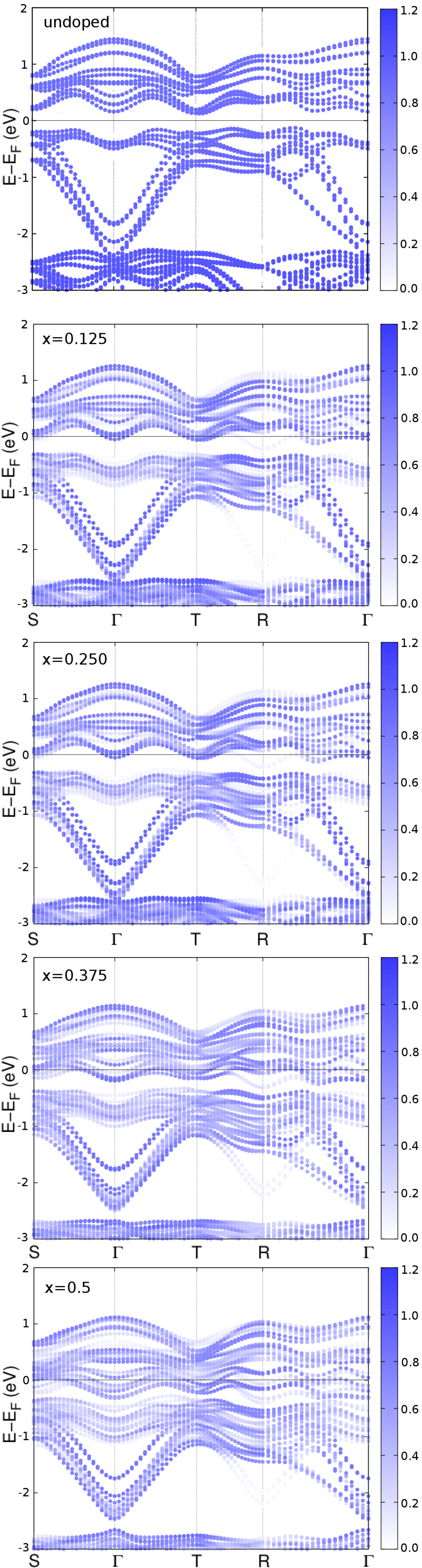}
\caption{Evolution of the band structure as a function of $x$ unfolded in the primitive cell.
The lateral bar indicates the amount of Bloch character given by Eq.~\ref{eq:bloch}.}
\label{fig:bands}
\end{figure}

This hypothesis is further verified by computing the electronic dispersion relation as a function of doping. The effective band structure, unfolded in the primitive cell, is shown in Fig.~\ref{fig:bands} for $x$=0, 0.125, 0.25, 0.375 and 0.5. The sharpness of the electronic bands measures the amount of 
Bloch character, as defined in Eq.~\ref{eq:bloch}, which accounts for the effect of the chemical disorder in the primitive cell.
At $x=0$ [Fig.~\ref{fig:bands}(a)], the system exhibits the characteristic insulating gap, and the effective band structure shows sharp bands, with a one-to-one correspondence between the primitive cell and the supercell.
Upon doping [Figs.~\ref{fig:bands}(b--e)], the energy bands appear with different degrees of intensity, due to the chemical disorder perturbing the electronic eigenstates.
Two different regimes can be recognized. For $0.125{\le}x{\le}0.375$ electron-doped the lower portion of the conduction band crosses the Fermi level and a feeble band forms around $R$ which approaches the valence band with increasing doping. This state can be interpreted as a bad metal regime as conductivity arises only from the the electron pockets mostly localized around the high-symmetry points separated by a small pseudogap (of the order of tens of meV) from the underlying valence band. A complete metal state is fully developed only at $x=0.5$ when the two $t_{2g}$ manifolds starts to overlap [Fig.~\ref{fig:bands}(b-d)].

Based on the above results we can therefore conclude, that upon Na$\rightarrow$Mg substitution NaOsO$_3$ undergoes a coupled electronic and magnetic transition in which the excess electrons move the system closer and closer to a itinerant metallic limit. Across the transition the local moments are continuously quenched and the pseudogap separating the valence and conduction bands is gradually reduced and finally closed for $x=0.5$. In the metallic state the moment is reduced by 30\% and should be most likely subjected to significant fluctuations.

This doping-induced MIT shares similarities with the temperature-driven MIT, in particular for what concerns the important role played by the attenuation of the local moment.  The analogy is not complete, however, since chemical doping does alter not rigidly the topology of the bands as it would be expected in a pure Lifshitz scenario. In particular, this difference emerges near the $R$ point, where the bandgap closes.

\begin{figure}[h]
\includegraphics[width=0.49\textwidth]{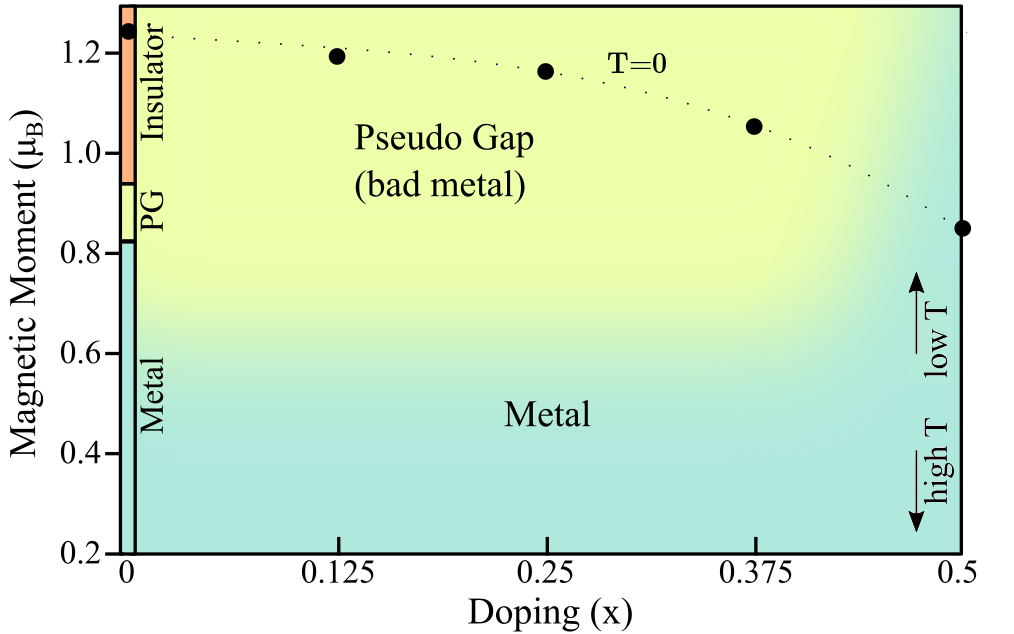}           
\caption{Schematic phase diagram of the MIT in NaOsO$_3$, showing the interplay between doping, magnetism and temperature. There different regimes as seen: (i) AFM insulator (low-T, $x=0$); (ii) bad-metal (low-T, $x\le{0.375}$) and (iii) metal (high-T any doping, $x=0.5$ any T). The circles represent the self-consistent value of the local spin moment at a given doping level computed at T=0.
}
\label{fig:pd}
\end{figure}

We conclude by linking the temperature to the doping induced MIT, by constructing a generalized phase diagram which rationalize within a unique picture the MIT in NaOsO$_3$. 
To this end we conducted a series of constrained magnetic moment calculations at each doping level to check how the electronic ground state changes as a function of the local moment~\cite{Kim2016, Liu2015}.
As already mentioned the T-driven MIT can be rationalized in terms of continuous damping of the local spin moment (associated with longitudinal and rotational spin fluctuations). This is schematically summarized on the left axis of Fig.~\ref{fig:pd} which shows the transition for the low-T AFM state to the paramagnetic high-T metal state through an intermediate pseudogap bad metal state.

When doping is considered only two regimes remain, metal and bad metal, and the boundary between one and the other is determined by the size of the local moment. Specifically, the self-consistent ground state solution (filled circles) follows the trend described by the evolution of the band structures (Fig.~\ref{fig:bands}): at low-T electron doping yields an insulator to bad-metal (pseudo gap) transition and at high doping the system enters the metallic state. Our data show that the transition between the pseudogap and metal state is controlled by the size of the local magnetic moment: by decreasing the magnitude of the local moment, by means of magnetically constrained DFT, NaOsO$_3$ gradually shifts from the pseudogap to the metal regime. In analogy with the undoped case we speculate that the strength of the local moment can be controlled by temperature effects. In fact, according to the Mohn and Wohlfarth approximation which assumes a linear temperature dependence of the local mean square moment amplitude~\cite{Mohn}
one can qualitatively map the magnetic moment axis (Fig.~\ref{fig:pd}, left axis) onto a temperature scale (Fig.~\ref{fig:pd}, right axis). We expect that, in analogy with the undoped case, with increasing temperature the systems will approach the metal state for doping level $x<0.5$. As already mentioned, at high charge carries concentrations ($x\approx$0.5) NaOsO$_3$ is predicted to be a magnetically itinerant good conductor.

\section{Conclusion}

In conclusion, by means of first principles calculations we have inspected the possibility to induce and control an insulator-to-metal transition in NaOsO$_3$. By combining magnetically constrained DFT+$U$ and the band unfolding scheme we have shown that excess electrons destroy the magnetic Lifshitz state and, depending on the doping concentration, the system can be tuned from a poor to a good metal by changing the amplitude of the local magnetic moment. Similarly to the temperature-driven MIT in the undoped compound, we propose that the transition to the itinerant metallic state can be 
controlled by temperature effects. As a minimal synthesis of this study, we propose a general phase diagram of NaOsO$_3$ in which the boundaries between the insulating, pseudogap and metallic regimes are determined by the size of the local moment, temperature and doping. 

This study reveals once more the peculiar nature of NaOsO$_3$ in which the electronic and magnetic degrees of freedom are tightly connected within an intrinsically itinerant background. Further studies might be envisioned to possibly preserve the insulating nature in NaOsO$_3$ in the low-doping range by creating trapping centers or point defects which could immobilize the excess charge carriers, thus providing an additional channel to design novel functionalizations or construct novel type of quantum state of matter.

\section*{Acknowledgements}
C.F. dedicates this work to the memory of Sandro Massidda, esteemed mentor and dear friend,
whose teachings and smiles will continue to be a source of inspiration.
This work was supported by the Austrian Science Fund (FWF) within the SFB ViCoM (Grant No. F41),
and by the joint DST (Indian Department of Science and Technology)-FWF project INDOX (I1490-N19).
Supercomputing time on the Vienna Scientific cluster (VSC) is gratefully acknowledged.

\bibliographystyle{apsrev4-1}
\bibliography{reference} 

\begin{thebibliography}{45}%
\makeatletter
\providecommand \@ifxundefined [1]{%
 \@ifx{#1\undefined}
}%
\providecommand \@ifnum [1]{%
 \ifnum #1\expandafter \@firstoftwo
 \else \expandafter \@secondoftwo
 \fi
}%
\providecommand \@ifx [1]{%
 \ifx #1\expandafter \@firstoftwo
 \else \expandafter \@secondoftwo
 \fi
}%
\providecommand \natexlab [1]{#1}%
\providecommand \enquote  [1]{``#1''}%
\providecommand \bibnamefont  [1]{#1}%
\providecommand \bibfnamefont [1]{#1}%
\providecommand \citenamefont [1]{#1}%
\providecommand \href@noop [0]{\@secondoftwo}%
\providecommand \href [0]{\begingroup \@sanitize@url \@href}%
\providecommand \@href[1]{\@@startlink{#1}\@@href}%
\providecommand \@@href[1]{\endgroup#1\@@endlink}%
\providecommand \@sanitize@url [0]{\catcode `\\12\catcode `\$12\catcode
  `\&12\catcode `\#12\catcode `\^12\catcode `\_12\catcode `\%12\relax}%
\providecommand \@@startlink[1]{}%
\providecommand \@@endlink[0]{}%
\providecommand \url  [0]{\begingroup\@sanitize@url \@url }%
\providecommand \@url [1]{\endgroup\@href {#1}{\urlprefix }}%
\providecommand \urlprefix  [0]{URL }%
\providecommand \Eprint [0]{\href }%
\providecommand \doibase [0]{http://dx.doi.org/}%
\providecommand \selectlanguage [0]{\@gobble}%
\providecommand \bibinfo  [0]{\@secondoftwo}%
\providecommand \bibfield  [0]{\@secondoftwo}%
\providecommand \translation [1]{[#1]}%
\providecommand \BibitemOpen [0]{}%
\providecommand \bibitemStop [0]{}%
\providecommand \bibitemNoStop [0]{.\EOS\space}%
\providecommand \EOS [0]{\spacefactor3000\relax}%
\providecommand \BibitemShut  [1]{\csname bibitem#1\endcsname}%
\let\auto@bib@innerbib\@empty
\bibitem [{\citenamefont {Imada}\ \emph {et~al.}(1998)\citenamefont {Imada},
  \citenamefont {Fujimori},\ and\ \citenamefont {Tokura}}]{Imada1998}%
  \BibitemOpen
  \bibfield  {author} {\bibinfo {author} {\bibfnamefont {M.}~\bibnamefont
  {Imada}}, \bibinfo {author} {\bibfnamefont {A.}~\bibnamefont {Fujimori}}, \
  and\ \bibinfo {author} {\bibfnamefont {Y.}~\bibnamefont {Tokura}},\ }\href
  {\doibase 10.1103/RevModPhys.70.1039} {\bibfield  {journal} {\bibinfo
  {journal} {Rev. Mod. Phys.}\ }\textbf {\bibinfo {volume} {70}},\ \bibinfo
  {pages} {1039} (\bibinfo {year} {1998})}\BibitemShut {NoStop}%
\bibitem [{\citenamefont {MOTT}(1968)}]{Mott1968}%
  \BibitemOpen
  \bibfield  {author} {\bibinfo {author} {\bibfnamefont {N.~F.}\ \bibnamefont
  {MOTT}},\ }\href {\doibase 10.1103/RevModPhys.40.677} {\bibfield  {journal}
  {\bibinfo  {journal} {Rev. Mod. Phys.}\ }\textbf {\bibinfo {volume} {40}},\
  \bibinfo {pages} {677} (\bibinfo {year} {1968})}\BibitemShut {NoStop}%
\bibitem [{\citenamefont {Yang}\ \emph {et~al.}(2011)\citenamefont {Yang},
  \citenamefont {Ko},\ and\ \citenamefont {Ramanathan}}]{Yang2011}%
  \BibitemOpen
  \bibfield  {author} {\bibinfo {author} {\bibfnamefont {Z.}~\bibnamefont
  {Yang}}, \bibinfo {author} {\bibfnamefont {C.}~\bibnamefont {Ko}}, \ and\
  \bibinfo {author} {\bibfnamefont {S.}~\bibnamefont {Ramanathan}},\ }\href
  {\doibase 10.1146/annurev-matsci-062910-100347} {\bibfield  {journal}
  {\bibinfo  {journal} {Annual Review of Materials Research}\ }\textbf
  {\bibinfo {volume} {41}},\ \bibinfo {pages} {337} (\bibinfo {year} {2011})},\
  \Eprint
  {http://arxiv.org/abs/https://doi.org/10.1146/annurev-matsci-062910-100347}
  {https://doi.org/10.1146/annurev-matsci-062910-100347} \BibitemShut {NoStop}%
\bibitem [{Wil(1931{\natexlab{a}})}]{Wilson458}%
  \BibitemOpen
  \href {\doibase 10.1098/rspa.1931.0162} {\bibfield  {journal} {\bibinfo
  {journal} {Proceedings of the Royal Society of London A: Mathematical,
  Physical and Engineering Sciences}\ }\textbf {\bibinfo {volume} {133}},\
  \bibinfo {pages} {458} (\bibinfo {year} {1931}{\natexlab{a}})},\ \Eprint
  {http://arxiv.org/abs/http://rspa.royalsocietypublishing.org/content/133/822/458.full.pdf}
  {http://rspa.royalsocietypublishing.org/content/133/822/458.full.pdf}
  \BibitemShut {NoStop}%
\bibitem [{Wil(1931{\natexlab{b}})}]{Wilson277}%
  \BibitemOpen
  \href {\doibase 10.1098/rspa.1931.0196} {\bibfield  {journal} {\bibinfo
  {journal} {Proceedings of the Royal Society of London A: Mathematical,
  Physical and Engineering Sciences}\ }\textbf {\bibinfo {volume} {134}},\
  \bibinfo {pages} {277} (\bibinfo {year} {1931}{\natexlab{b}})},\ \Eprint
  {http://arxiv.org/abs/http://rspa.royalsocietypublishing.org/content/134/823/277.full.pdf}
  {http://rspa.royalsocietypublishing.org/content/134/823/277.full.pdf}
  \BibitemShut {NoStop}%
\bibitem [{\citenamefont {de~Boer}\ and\ \citenamefont {Verwey}(1937)}]{Boer}%
  \BibitemOpen
  \bibfield  {author} {\bibinfo {author} {\bibfnamefont {J.~H.}\ \bibnamefont
  {de~Boer}}\ and\ \bibinfo {author} {\bibfnamefont {E.~J.~W.}\ \bibnamefont
  {Verwey}},\ }\href@noop {} {\bibfield  {journal} {\bibinfo  {journal}
  {Proceedings of the Physical Society}\ }\textbf {\bibinfo {volume} {49}},\
  \bibinfo {pages} {59} (\bibinfo {year} {1937})}\BibitemShut {NoStop}%
\bibitem [{\citenamefont {Mott}(1949)}]{Mott1949}%
  \BibitemOpen
  \bibfield  {author} {\bibinfo {author} {\bibfnamefont {N.~F.}\ \bibnamefont
  {Mott}},\ }\href@noop {} {\bibfield  {journal} {\bibinfo  {journal}
  {Proceedings of the Physical Society. Section A}\ }\textbf {\bibinfo {volume}
  {62}},\ \bibinfo {pages} {416} (\bibinfo {year} {1949})}\BibitemShut
  {NoStop}%
\bibitem [{\citenamefont {Kim}\ \emph {et~al.}(2008)\citenamefont {Kim},
  \citenamefont {Jin}, \citenamefont {Moon}, \citenamefont {Kim}, \citenamefont
  {Park}, \citenamefont {Leem}, \citenamefont {Yu}, \citenamefont {Noh},
  \citenamefont {Kim}, \citenamefont {Oh}, \citenamefont {Park}, \citenamefont
  {Durairaj}, \citenamefont {Cao},\ and\ \citenamefont {Rotenberg}}]{Kim2008}%
  \BibitemOpen
  \bibfield  {author} {\bibinfo {author} {\bibfnamefont {B.~J.}\ \bibnamefont
  {Kim}}, \bibinfo {author} {\bibfnamefont {H.}~\bibnamefont {Jin}}, \bibinfo
  {author} {\bibfnamefont {S.}~\bibnamefont {Moon}}, \bibinfo {author}
  {\bibfnamefont {J.~Y.}\ \bibnamefont {Kim}}, \bibinfo {author} {\bibfnamefont
  {B.~G.}\ \bibnamefont {Park}}, \bibinfo {author} {\bibfnamefont
  {C.}~\bibnamefont {Leem}}, \bibinfo {author} {\bibfnamefont {J.}~\bibnamefont
  {Yu}}, \bibinfo {author} {\bibfnamefont {T.}~\bibnamefont {Noh}}, \bibinfo
  {author} {\bibfnamefont {C.}~\bibnamefont {Kim}}, \bibinfo {author}
  {\bibfnamefont {S.~J.}\ \bibnamefont {Oh}}, \bibinfo {author} {\bibfnamefont
  {J.~H.}\ \bibnamefont {Park}}, \bibinfo {author} {\bibfnamefont
  {V.}~\bibnamefont {Durairaj}}, \bibinfo {author} {\bibfnamefont
  {G.}~\bibnamefont {Cao}}, \ and\ \bibinfo {author} {\bibfnamefont
  {E.}~\bibnamefont {Rotenberg}},\ }\href@noop {} {\bibfield  {journal}
  {\bibinfo  {journal} {Phys. Rev. Lett.}\ }\textbf {\bibinfo {volume} {101}},\
  \bibinfo {pages} {076402} (\bibinfo {year} {2008})}\BibitemShut {NoStop}%
\bibitem [{\citenamefont {Kim}\ \emph {et~al.}(2009)\citenamefont {Kim},
  \citenamefont {Ohsumi}, \citenamefont {Komesu}, \citenamefont {Sakai},
  \citenamefont {Morita}, \citenamefont {Takagi},\ and\ \citenamefont
  {Arima}}]{Kim2009}%
  \BibitemOpen
  \bibfield  {author} {\bibinfo {author} {\bibfnamefont {B.~J.}\ \bibnamefont
  {Kim}}, \bibinfo {author} {\bibfnamefont {H.}~\bibnamefont {Ohsumi}},
  \bibinfo {author} {\bibfnamefont {T.}~\bibnamefont {Komesu}}, \bibinfo
  {author} {\bibfnamefont {S.}~\bibnamefont {Sakai}}, \bibinfo {author}
  {\bibfnamefont {T.}~\bibnamefont {Morita}}, \bibinfo {author} {\bibfnamefont
  {H.}~\bibnamefont {Takagi}}, \ and\ \bibinfo {author} {\bibfnamefont
  {T.}~\bibnamefont {Arima}},\ }\href@noop {} {\bibfield  {journal} {\bibinfo
  {journal} {Science}\ }\textbf {\bibinfo {volume} {323}},\ \bibinfo {pages}
  {1329} (\bibinfo {year} {2009})}\BibitemShut {NoStop}%
\bibitem [{\citenamefont {Jackeli}\ and\ \citenamefont
  {Khaliullin}(2009)}]{Jackeli2009}%
  \BibitemOpen
  \bibfield  {author} {\bibinfo {author} {\bibfnamefont {G.}~\bibnamefont
  {Jackeli}}\ and\ \bibinfo {author} {\bibfnamefont {G.}~\bibnamefont
  {Khaliullin}},\ }\href@noop {} {\bibfield  {journal} {\bibinfo  {journal}
  {Phys. Rev. Lett.}\ }\textbf {\bibinfo {volume} {102}},\ \bibinfo {pages}
  {017205} (\bibinfo {year} {2009})}\BibitemShut {NoStop}%
\bibitem [{\citenamefont {Liu}\ \emph {et~al.}(2015)\citenamefont {Liu},
  \citenamefont {Khmelevskyi}, \citenamefont {Kim}, \citenamefont {Marsman},
  \citenamefont {Li}, \citenamefont {Chen}, \citenamefont {Sarma},
  \citenamefont {Kresse},\ and\ \citenamefont {Franchini}}]{Liu2015}%
  \BibitemOpen
  \bibfield  {author} {\bibinfo {author} {\bibfnamefont {P.}~\bibnamefont
  {Liu}}, \bibinfo {author} {\bibfnamefont {S.}~\bibnamefont {Khmelevskyi}},
  \bibinfo {author} {\bibfnamefont {B.}~\bibnamefont {Kim}}, \bibinfo {author}
  {\bibfnamefont {M.}~\bibnamefont {Marsman}}, \bibinfo {author} {\bibfnamefont
  {D.}~\bibnamefont {Li}}, \bibinfo {author} {\bibfnamefont {X.-Q.}\
  \bibnamefont {Chen}}, \bibinfo {author} {\bibfnamefont {D.~D.}\ \bibnamefont
  {Sarma}}, \bibinfo {author} {\bibfnamefont {G.}~\bibnamefont {Kresse}}, \
  and\ \bibinfo {author} {\bibfnamefont {C.}~\bibnamefont {Franchini}},\
  }\href@noop {} {\bibfield  {journal} {\bibinfo  {journal} {Phys. Rev. B}\
  }\textbf {\bibinfo {volume} {92}},\ \bibinfo {pages} {054428} (\bibinfo
  {year} {2015})}\BibitemShut {NoStop}%
\bibitem [{\citenamefont {Liu}\ \emph {et~al.}(2018)\citenamefont {Liu},
  \citenamefont {Kim}, \citenamefont {Chen}, \citenamefont {Sarma},
  \citenamefont {Kresse},\ and\ \citenamefont {Franchini}}]{Liu2018}%
  \BibitemOpen
  \bibfield  {author} {\bibinfo {author} {\bibfnamefont {P.}~\bibnamefont
  {Liu}}, \bibinfo {author} {\bibfnamefont {B.}~\bibnamefont {Kim}}, \bibinfo
  {author} {\bibfnamefont {X.-Q.}\ \bibnamefont {Chen}}, \bibinfo {author}
  {\bibfnamefont {D.~D.}\ \bibnamefont {Sarma}}, \bibinfo {author}
  {\bibfnamefont {G.}~\bibnamefont {Kresse}}, \ and\ \bibinfo {author}
  {\bibfnamefont {C.}~\bibnamefont {Franchini}},\ }\href {\doibase
  10.1103/PhysRevMaterials.2.075003} {\bibfield  {journal} {\bibinfo  {journal}
  {Phys. Rev. Materials}\ }\textbf {\bibinfo {volume} {2}},\ \bibinfo {pages}
  {075003} (\bibinfo {year} {2018})}\BibitemShut {NoStop}%
\bibitem [{\citenamefont {Calder}\ \emph {et~al.}(2012)\citenamefont {Calder},
  \citenamefont {Garlea}, \citenamefont {McMorrow}, \citenamefont {Lumsden},
  \citenamefont {Stone}, \citenamefont {Lang}, \citenamefont {Kim},
  \citenamefont {Schlueter}, \citenamefont {Shi}, \citenamefont {Yamaura},
  \citenamefont {Sun}, \citenamefont {Tsujimoto},\ and\ \citenamefont
  {Christianson}}]{Calder2012}%
  \BibitemOpen
  \bibfield  {author} {\bibinfo {author} {\bibfnamefont {S.}~\bibnamefont
  {Calder}}, \bibinfo {author} {\bibfnamefont {V.~O.}\ \bibnamefont {Garlea}},
  \bibinfo {author} {\bibfnamefont {D.~F.}\ \bibnamefont {McMorrow}}, \bibinfo
  {author} {\bibfnamefont {M.~D.}\ \bibnamefont {Lumsden}}, \bibinfo {author}
  {\bibfnamefont {M.~B.}\ \bibnamefont {Stone}}, \bibinfo {author}
  {\bibfnamefont {J.~C.}\ \bibnamefont {Lang}}, \bibinfo {author}
  {\bibfnamefont {J.-W.}\ \bibnamefont {Kim}}, \bibinfo {author} {\bibfnamefont
  {J.~A.}\ \bibnamefont {Schlueter}}, \bibinfo {author} {\bibfnamefont {Y.~G.}\
  \bibnamefont {Shi}}, \bibinfo {author} {\bibfnamefont {K.}~\bibnamefont
  {Yamaura}}, \bibinfo {author} {\bibfnamefont {Y.~S.}\ \bibnamefont {Sun}},
  \bibinfo {author} {\bibfnamefont {Y.}~\bibnamefont {Tsujimoto}}, \ and\
  \bibinfo {author} {\bibfnamefont {A.~D.}\ \bibnamefont {Christianson}},\
  }\href {\doibase 10.1103/PhysRevLett.108.257209} {\bibfield  {journal}
  {\bibinfo  {journal} {Phys. Rev. Lett.}\ }\textbf {\bibinfo {volume} {108}},\
  \bibinfo {pages} {257209} (\bibinfo {year} {2012})}\BibitemShut {NoStop}%
\bibitem [{\citenamefont {Jung}\ \emph {et~al.}(2013)\citenamefont {Jung},
  \citenamefont {Song}, \citenamefont {Lee},\ and\ \citenamefont
  {Pickett}}]{Jung2013}%
  \BibitemOpen
  \bibfield  {author} {\bibinfo {author} {\bibfnamefont {M.-C.}\ \bibnamefont
  {Jung}}, \bibinfo {author} {\bibfnamefont {Y.-J.}\ \bibnamefont {Song}},
  \bibinfo {author} {\bibfnamefont {K.-W.}\ \bibnamefont {Lee}}, \ and\
  \bibinfo {author} {\bibfnamefont {W.~E.}\ \bibnamefont {Pickett}},\ }\href
  {\doibase 10.1103/PhysRevB.87.115119} {\bibfield  {journal} {\bibinfo
  {journal} {Phys. Rev. B}\ }\textbf {\bibinfo {volume} {87}},\ \bibinfo
  {pages} {115119} (\bibinfo {year} {2013})}\BibitemShut {NoStop}%
\bibitem [{\citenamefont {Middey}\ \emph {et~al.}(2014)\citenamefont {Middey},
  \citenamefont {Debnath}, \citenamefont {Mahadevan},\ and\ \citenamefont
  {Sarma}}]{Middey2014}%
  \BibitemOpen
  \bibfield  {author} {\bibinfo {author} {\bibfnamefont {S.}~\bibnamefont
  {Middey}}, \bibinfo {author} {\bibfnamefont {S.}~\bibnamefont {Debnath}},
  \bibinfo {author} {\bibfnamefont {P.}~\bibnamefont {Mahadevan}}, \ and\
  \bibinfo {author} {\bibfnamefont {D.~D.}\ \bibnamefont {Sarma}},\ }\href
  {\doibase 10.1103/PhysRevB.89.134416} {\bibfield  {journal} {\bibinfo
  {journal} {Phys. Rev. B}\ }\textbf {\bibinfo {volume} {89}},\ \bibinfo
  {pages} {134416} (\bibinfo {year} {2014})}\BibitemShut {NoStop}%
\bibitem [{\citenamefont {Kim}\ \emph {et~al.}(2016)\citenamefont {Kim},
  \citenamefont {Liu}, \citenamefont {Erg\"onenc}, \citenamefont {Toschi},
  \citenamefont {Khmelevskyi},\ and\ \citenamefont {Franchini}}]{Kim2016}%
  \BibitemOpen
  \bibfield  {author} {\bibinfo {author} {\bibfnamefont {B.}~\bibnamefont
  {Kim}}, \bibinfo {author} {\bibfnamefont {P.}~\bibnamefont {Liu}}, \bibinfo
  {author} {\bibfnamefont {Z.}~\bibnamefont {Erg\"onenc}}, \bibinfo {author}
  {\bibfnamefont {A.}~\bibnamefont {Toschi}}, \bibinfo {author} {\bibfnamefont
  {S.}~\bibnamefont {Khmelevskyi}}, \ and\ \bibinfo {author} {\bibfnamefont
  {C.}~\bibnamefont {Franchini}},\ }\href {\doibase 10.1103/PhysRevB.94.241113}
  {\bibfield  {journal} {\bibinfo  {journal} {Phys. Rev. B}\ }\textbf {\bibinfo
  {volume} {94}},\ \bibinfo {pages} {241113} (\bibinfo {year}
  {2016})}\BibitemShut {NoStop}%
\bibitem [{\citenamefont {Mandrus}\ \emph {et~al.}(2001)\citenamefont
  {Mandrus}, \citenamefont {Thompson}, \citenamefont {Gaal}, \citenamefont
  {Forro}, \citenamefont {Bryan}, \citenamefont {Chakoumakos}, \citenamefont
  {Woods}, \citenamefont {Sales}, \citenamefont {Fishman},\ and\ \citenamefont
  {Keppens}}]{Mandrus2001}%
  \BibitemOpen
  \bibfield  {author} {\bibinfo {author} {\bibfnamefont {D.}~\bibnamefont
  {Mandrus}}, \bibinfo {author} {\bibfnamefont {J.~R.}\ \bibnamefont
  {Thompson}}, \bibinfo {author} {\bibfnamefont {R.}~\bibnamefont {Gaal}},
  \bibinfo {author} {\bibfnamefont {L.}~\bibnamefont {Forro}}, \bibinfo
  {author} {\bibfnamefont {J.~C.}\ \bibnamefont {Bryan}}, \bibinfo {author}
  {\bibfnamefont {B.~C.}\ \bibnamefont {Chakoumakos}}, \bibinfo {author}
  {\bibfnamefont {L.~M.}\ \bibnamefont {Woods}}, \bibinfo {author}
  {\bibfnamefont {B.~C.}\ \bibnamefont {Sales}}, \bibinfo {author}
  {\bibfnamefont {R.~S.}\ \bibnamefont {Fishman}}, \ and\ \bibinfo {author}
  {\bibfnamefont {V.}~\bibnamefont {Keppens}},\ }\href {\doibase
  10.1103/PhysRevB.63.195104} {\bibfield  {journal} {\bibinfo  {journal} {Phys.
  Rev. B}\ }\textbf {\bibinfo {volume} {63}},\ \bibinfo {pages} {195104}
  (\bibinfo {year} {2001})}\BibitemShut {NoStop}%
\bibitem [{\citenamefont {Yamaura}\ \emph {et~al.}(2012)\citenamefont
  {Yamaura}, \citenamefont {Ohgushi}, \citenamefont {Ohsumi}, \citenamefont
  {Hasegawa}, \citenamefont {Yamauchi}, \citenamefont {Sugimoto}, \citenamefont
  {Takeshita}, \citenamefont {Tokuda}, \citenamefont {Takata}, \citenamefont
  {Udagawa}, \citenamefont {Takigawa}, \citenamefont {Harima}, \citenamefont
  {Arima},\ and\ \citenamefont {Hiroi}}]{Yamaura2012}%
  \BibitemOpen
  \bibfield  {author} {\bibinfo {author} {\bibfnamefont {J.}~\bibnamefont
  {Yamaura}}, \bibinfo {author} {\bibfnamefont {K.}~\bibnamefont {Ohgushi}},
  \bibinfo {author} {\bibfnamefont {H.}~\bibnamefont {Ohsumi}}, \bibinfo
  {author} {\bibfnamefont {T.}~\bibnamefont {Hasegawa}}, \bibinfo {author}
  {\bibfnamefont {I.}~\bibnamefont {Yamauchi}}, \bibinfo {author}
  {\bibfnamefont {K.}~\bibnamefont {Sugimoto}}, \bibinfo {author}
  {\bibfnamefont {S.}~\bibnamefont {Takeshita}}, \bibinfo {author}
  {\bibfnamefont {A.}~\bibnamefont {Tokuda}}, \bibinfo {author} {\bibfnamefont
  {M.}~\bibnamefont {Takata}}, \bibinfo {author} {\bibfnamefont
  {M.}~\bibnamefont {Udagawa}}, \bibinfo {author} {\bibfnamefont
  {M.}~\bibnamefont {Takigawa}}, \bibinfo {author} {\bibfnamefont
  {H.}~\bibnamefont {Harima}}, \bibinfo {author} {\bibfnamefont
  {T.}~\bibnamefont {Arima}}, \ and\ \bibinfo {author} {\bibfnamefont
  {Z.}~\bibnamefont {Hiroi}},\ }\href {\doibase 10.1103/PhysRevLett.108.247205}
  {\bibfield  {journal} {\bibinfo  {journal} {Phys. Rev. Lett.}\ }\textbf
  {\bibinfo {volume} {108}},\ \bibinfo {pages} {247205} (\bibinfo {year}
  {2012})}\BibitemShut {NoStop}%
\bibitem [{\citenamefont {Hiroi}\ \emph {et~al.}(2015)\citenamefont {Hiroi},
  \citenamefont {Yamaura}, \citenamefont {Hirose}, \citenamefont {Nagashima},\
  and\ \citenamefont {Okamoto}}]{Hiroi2015}%
  \BibitemOpen
  \bibfield  {author} {\bibinfo {author} {\bibfnamefont {Z.}~\bibnamefont
  {Hiroi}}, \bibinfo {author} {\bibfnamefont {J.}~\bibnamefont {Yamaura}},
  \bibinfo {author} {\bibfnamefont {T.}~\bibnamefont {Hirose}}, \bibinfo
  {author} {\bibfnamefont {I.}~\bibnamefont {Nagashima}}, \ and\ \bibinfo
  {author} {\bibfnamefont {Y.}~\bibnamefont {Okamoto}},\ }\href {\doibase
  10.1063/1.4907734} {\bibfield  {journal} {\bibinfo  {journal} {APL
  Materials}\ }\textbf {\bibinfo {volume} {3}},\ \bibinfo {pages} {041501}
  (\bibinfo {year} {2015})},\ \Eprint
  {http://arxiv.org/abs/https://doi.org/10.1063/1.4907734}
  {https://doi.org/10.1063/1.4907734} \BibitemShut {NoStop}%
\bibitem [{\citenamefont {Padilla}\ \emph {et~al.}(2002)\citenamefont
  {Padilla}, \citenamefont {Mandrus},\ and\ \citenamefont
  {Basov}}]{Padilla2002}%
  \BibitemOpen
  \bibfield  {author} {\bibinfo {author} {\bibfnamefont {W.~J.}\ \bibnamefont
  {Padilla}}, \bibinfo {author} {\bibfnamefont {D.}~\bibnamefont {Mandrus}}, \
  and\ \bibinfo {author} {\bibfnamefont {D.~N.}\ \bibnamefont {Basov}},\ }\href
  {\doibase 10.1103/PhysRevB.66.035120} {\bibfield  {journal} {\bibinfo
  {journal} {Phys. Rev. B}\ }\textbf {\bibinfo {volume} {66}},\ \bibinfo
  {pages} {035120} (\bibinfo {year} {2002})}\BibitemShut {NoStop}%
\bibitem [{\citenamefont {Vecchio}\ \emph {et~al.}(2013)\citenamefont
  {Vecchio}, \citenamefont {Perucchi}, \citenamefont {Di~Pietro}, \citenamefont
  {Limaj}, \citenamefont {Schade}, \citenamefont {Sun}, \citenamefont {Arai},
  \citenamefont {Yamaura},\ and\ \citenamefont {Lupi}}]{Vecchio2013}%
  \BibitemOpen
  \bibfield  {author} {\bibinfo {author} {\bibfnamefont {I.~L.}\ \bibnamefont
  {Vecchio}}, \bibinfo {author} {\bibfnamefont {A.}~\bibnamefont {Perucchi}},
  \bibinfo {author} {\bibfnamefont {P.}~\bibnamefont {Di~Pietro}}, \bibinfo
  {author} {\bibfnamefont {O.}~\bibnamefont {Limaj}}, \bibinfo {author}
  {\bibfnamefont {U.}~\bibnamefont {Schade}}, \bibinfo {author} {\bibfnamefont
  {Y.}~\bibnamefont {Sun}}, \bibinfo {author} {\bibfnamefont {M.}~\bibnamefont
  {Arai}}, \bibinfo {author} {\bibfnamefont {K.}~\bibnamefont {Yamaura}}, \
  and\ \bibinfo {author} {\bibfnamefont {S.}~\bibnamefont {Lupi}},\ }\href
  {https://doi.org/10.1038/srep02990} {\bibfield  {journal} {\bibinfo
  {journal} {Scientific Reports}\ }\textbf {\bibinfo {volume} {3}},\ \bibinfo
  {pages} {2990 EP } (\bibinfo {year} {2013})},\ \bibinfo {note}
  {article}\BibitemShut {NoStop}%
\bibitem [{\citenamefont {Slater}(1951)}]{Slater}%
  \BibitemOpen
  \bibfield  {author} {\bibinfo {author} {\bibfnamefont {J.~C.}\ \bibnamefont
  {Slater}},\ }\href {\doibase 10.1103/PhysRev.82.538} {\bibfield  {journal}
  {\bibinfo  {journal} {Phys. Rev.}\ }\textbf {\bibinfo {volume} {82}},\
  \bibinfo {pages} {538} (\bibinfo {year} {1951})}\BibitemShut {NoStop}%
\bibitem [{\citenamefont {Lifshitz}(1960)}]{Lifshitz}%
  \BibitemOpen
  \bibfield  {author} {\bibinfo {author} {\bibfnamefont {M.}~\bibnamefont
  {Lifshitz}},\ }\href@noop {} {\bibfield  {journal} {\bibinfo  {journal} {Sov.
  Phys. JETP}\ }\textbf {\bibinfo {volume} {11}},\ \bibinfo {pages} {1130}
  (\bibinfo {year} {1960})}\BibitemShut {NoStop}%
\bibitem [{\citenamefont {Shi}\ \emph {et~al.}(2009)\citenamefont {Shi},
  \citenamefont {Guo}, \citenamefont {Yu}, \citenamefont {Arai}, \citenamefont
  {Belik}, \citenamefont {Sato}, \citenamefont {Yamaura}, \citenamefont
  {Takayama-Muromachi}, \citenamefont {Tian}, \citenamefont {Yang},
  \citenamefont {Li}, \citenamefont {Varga}, \citenamefont {Mitchell},\ and\
  \citenamefont {Okamoto}}]{Shi2009}%
  \BibitemOpen
  \bibfield  {author} {\bibinfo {author} {\bibfnamefont {Y.~G.}\ \bibnamefont
  {Shi}}, \bibinfo {author} {\bibfnamefont {Y.~F.}\ \bibnamefont {Guo}},
  \bibinfo {author} {\bibfnamefont {S.}~\bibnamefont {Yu}}, \bibinfo {author}
  {\bibfnamefont {M.}~\bibnamefont {Arai}}, \bibinfo {author} {\bibfnamefont
  {A.~A.}\ \bibnamefont {Belik}}, \bibinfo {author} {\bibfnamefont
  {A.}~\bibnamefont {Sato}}, \bibinfo {author} {\bibfnamefont {K.}~\bibnamefont
  {Yamaura}}, \bibinfo {author} {\bibfnamefont {E.}~\bibnamefont
  {Takayama-Muromachi}}, \bibinfo {author} {\bibfnamefont {H.~F.}\ \bibnamefont
  {Tian}}, \bibinfo {author} {\bibfnamefont {H.~X.}\ \bibnamefont {Yang}},
  \bibinfo {author} {\bibfnamefont {J.~Q.}\ \bibnamefont {Li}}, \bibinfo
  {author} {\bibfnamefont {T.}~\bibnamefont {Varga}}, \bibinfo {author}
  {\bibfnamefont {J.~F.}\ \bibnamefont {Mitchell}}, \ and\ \bibinfo {author}
  {\bibfnamefont {S.}~\bibnamefont {Okamoto}},\ }\href {\doibase
  10.1103/PhysRevB.80.161104} {\bibfield  {journal} {\bibinfo  {journal} {Phys.
  Rev. B}\ }\textbf {\bibinfo {volume} {80}},\ \bibinfo {pages} {161104}
  (\bibinfo {year} {2009})}\BibitemShut {NoStop}%
\bibitem [{\citenamefont {Calder}\ \emph {et~al.}(2017)\citenamefont {Calder},
  \citenamefont {Vale}, \citenamefont {Bogdanov}, \citenamefont {Donnerer},
  \citenamefont {Pincini}, \citenamefont {Moretti~Sala}, \citenamefont {Liu},
  \citenamefont {Upton}, \citenamefont {Casa}, \citenamefont {Shi},
  \citenamefont {Tsujimoto}, \citenamefont {Yamaura}, \citenamefont {Hill},
  \citenamefont {van~den Brink}, \citenamefont {McMorrow},\ and\ \citenamefont
  {Christianson}}]{Calder2017}%
  \BibitemOpen
  \bibfield  {author} {\bibinfo {author} {\bibfnamefont {S.}~\bibnamefont
  {Calder}}, \bibinfo {author} {\bibfnamefont {J.~G.}\ \bibnamefont {Vale}},
  \bibinfo {author} {\bibfnamefont {N.}~\bibnamefont {Bogdanov}}, \bibinfo
  {author} {\bibfnamefont {C.}~\bibnamefont {Donnerer}}, \bibinfo {author}
  {\bibfnamefont {D.}~\bibnamefont {Pincini}}, \bibinfo {author} {\bibfnamefont
  {M.}~\bibnamefont {Moretti~Sala}}, \bibinfo {author} {\bibfnamefont
  {X.}~\bibnamefont {Liu}}, \bibinfo {author} {\bibfnamefont {M.~H.}\
  \bibnamefont {Upton}}, \bibinfo {author} {\bibfnamefont {D.}~\bibnamefont
  {Casa}}, \bibinfo {author} {\bibfnamefont {Y.~G.}\ \bibnamefont {Shi}},
  \bibinfo {author} {\bibfnamefont {Y.}~\bibnamefont {Tsujimoto}}, \bibinfo
  {author} {\bibfnamefont {K.}~\bibnamefont {Yamaura}}, \bibinfo {author}
  {\bibfnamefont {J.~P.}\ \bibnamefont {Hill}}, \bibinfo {author}
  {\bibfnamefont {J.}~\bibnamefont {van~den Brink}}, \bibinfo {author}
  {\bibfnamefont {D.~F.}\ \bibnamefont {McMorrow}}, \ and\ \bibinfo {author}
  {\bibfnamefont {A.~D.}\ \bibnamefont {Christianson}},\ }\href {\doibase
  10.1103/PhysRevB.95.020413} {\bibfield  {journal} {\bibinfo  {journal} {Phys.
  Rev. B}\ }\textbf {\bibinfo {volume} {95}},\ \bibinfo {pages} {020413}
  (\bibinfo {year} {2017})}\BibitemShut {NoStop}%
\bibitem [{\citenamefont {Goodenough}(1968)}]{good}%
  \BibitemOpen
  \bibfield  {author} {\bibinfo {author} {\bibfnamefont {J.~B.}\ \bibnamefont
  {Goodenough}},\ }\href {\doibase 10.1103/PhysRev.171.466} {\bibfield
  {journal} {\bibinfo  {journal} {Phys. Rev.}\ }\textbf {\bibinfo {volume}
  {171}},\ \bibinfo {pages} {466} (\bibinfo {year} {1968})}\BibitemShut
  {NoStop}%
\bibitem [{\citenamefont {Singh}\ \emph {et~al.}(2018)\citenamefont {Singh},
  \citenamefont {Mohapatra}, \citenamefont {Bhandari},\ and\ \citenamefont
  {Satpathy}}]{Singh2018}%
  \BibitemOpen
  \bibfield  {author} {\bibinfo {author} {\bibfnamefont {A.}~\bibnamefont
  {Singh}}, \bibinfo {author} {\bibfnamefont {S.}~\bibnamefont {Mohapatra}},
  \bibinfo {author} {\bibfnamefont {C.}~\bibnamefont {Bhandari}}, \ and\
  \bibinfo {author} {\bibfnamefont {S.}~\bibnamefont {Satpathy}},\ }\href
  {http://stacks.iop.org/2399-6528/2/i=11/a=115016} {\bibfield  {journal}
  {\bibinfo  {journal} {Journal of Physics Communications}\ }\textbf {\bibinfo
  {volume} {2}},\ \bibinfo {pages} {115016} (\bibinfo {year}
  {2018})}\BibitemShut {NoStop}%
\bibitem [{\citenamefont {Liu}\ \emph {et~al.}(2016)\citenamefont {Liu},
  \citenamefont {Reticcioli}, \citenamefont {Kim}, \citenamefont {Continenza},
  \citenamefont {Kresse}, \citenamefont {Sarma}, \citenamefont {Chen},\ and\
  \citenamefont {Franchini}}]{Liu2016}%
  \BibitemOpen
  \bibfield  {author} {\bibinfo {author} {\bibfnamefont {P.}~\bibnamefont
  {Liu}}, \bibinfo {author} {\bibfnamefont {M.}~\bibnamefont {Reticcioli}},
  \bibinfo {author} {\bibfnamefont {B.}~\bibnamefont {Kim}}, \bibinfo {author}
  {\bibfnamefont {A.}~\bibnamefont {Continenza}}, \bibinfo {author}
  {\bibfnamefont {G.}~\bibnamefont {Kresse}}, \bibinfo {author} {\bibfnamefont
  {D.~D.}\ \bibnamefont {Sarma}}, \bibinfo {author} {\bibfnamefont {X.-Q.}\
  \bibnamefont {Chen}}, \ and\ \bibinfo {author} {\bibfnamefont
  {C.}~\bibnamefont {Franchini}},\ }\href {\doibase 10.1103/PhysRevB.94.195145}
  {\bibfield  {journal} {\bibinfo  {journal} {Phys. Rev. B}\ }\textbf {\bibinfo
  {volume} {94}},\ \bibinfo {pages} {195145} (\bibinfo {year}
  {2016})}\BibitemShut {NoStop}%
\bibitem [{\citenamefont {Franchini}\ \emph {et~al.}(2005)\citenamefont
  {Franchini}, \citenamefont {Sanna}, \citenamefont {Massidda},\ and\
  \citenamefont {Gauzzi}}]{Franchini2005}%
  \BibitemOpen
  \bibfield  {author} {\bibinfo {author} {\bibfnamefont {C.}~\bibnamefont
  {Franchini}}, \bibinfo {author} {\bibfnamefont {A.}~\bibnamefont {Sanna}},
  \bibinfo {author} {\bibfnamefont {S.}~\bibnamefont {Massidda}}, \ and\
  \bibinfo {author} {\bibfnamefont {A.}~\bibnamefont {Gauzzi}},\ }\href
  {http://stacks.iop.org/0295-5075/71/i=6/a=952} {\bibfield  {journal}
  {\bibinfo  {journal} {EPL (Europhysics Letters)}\ }\textbf {\bibinfo {volume}
  {71}},\ \bibinfo {pages} {952} (\bibinfo {year} {2005})}\BibitemShut
  {NoStop}%
\bibitem [{\citenamefont {Sanna}\ \emph {et~al.}(2004)\citenamefont {Sanna},
  \citenamefont {Franchini}, \citenamefont {Massidda},\ and\ \citenamefont
  {Gauzzi}}]{Sanna2004}%
  \BibitemOpen
  \bibfield  {author} {\bibinfo {author} {\bibfnamefont {A.}~\bibnamefont
  {Sanna}}, \bibinfo {author} {\bibfnamefont {C.}~\bibnamefont {Franchini}},
  \bibinfo {author} {\bibfnamefont {S.}~\bibnamefont {Massidda}}, \ and\
  \bibinfo {author} {\bibfnamefont {A.}~\bibnamefont {Gauzzi}},\ }\href
  {\doibase 10.1103/PhysRevB.70.235102} {\bibfield  {journal} {\bibinfo
  {journal} {Phys. Rev. B}\ }\textbf {\bibinfo {volume} {70}},\ \bibinfo
  {pages} {235102} (\bibinfo {year} {2004})}\BibitemShut {NoStop}%
\bibitem [{\citenamefont {He}\ \emph {et~al.}(2012)\citenamefont {He},
  \citenamefont {Chen}, \citenamefont {Chen},\ and\ \citenamefont
  {Franchini}}]{He2012}%
  \BibitemOpen
  \bibfield  {author} {\bibinfo {author} {\bibfnamefont {J.}~\bibnamefont
  {He}}, \bibinfo {author} {\bibfnamefont {M.-X.}\ \bibnamefont {Chen}},
  \bibinfo {author} {\bibfnamefont {X.-Q.}\ \bibnamefont {Chen}}, \ and\
  \bibinfo {author} {\bibfnamefont {C.}~\bibnamefont {Franchini}},\ }\href
  {\doibase 10.1103/PhysRevB.85.195135} {\bibfield  {journal} {\bibinfo
  {journal} {Phys. Rev. B}\ }\textbf {\bibinfo {volume} {85}},\ \bibinfo
  {pages} {195135} (\bibinfo {year} {2012})}\BibitemShut {NoStop}%
\bibitem [{\citenamefont {Kim}\ \emph {et~al.}(2017)\citenamefont {Kim},
  \citenamefont {Liu},\ and\ \citenamefont {Franchini}}]{Kim2017}%
  \BibitemOpen
  \bibfield  {author} {\bibinfo {author} {\bibfnamefont {B.}~\bibnamefont
  {Kim}}, \bibinfo {author} {\bibfnamefont {P.}~\bibnamefont {Liu}}, \ and\
  \bibinfo {author} {\bibfnamefont {C.}~\bibnamefont {Franchini}},\ }\href
  {\doibase 10.1103/PhysRevB.95.115111} {\bibfield  {journal} {\bibinfo
  {journal} {Phys. Rev. B}\ }\textbf {\bibinfo {volume} {95}},\ \bibinfo
  {pages} {115111} (\bibinfo {year} {2017})}\BibitemShut {NoStop}%
\bibitem [{\citenamefont {Bl\"ochl}(1994)}]{PhysRevB.50.17953}%
  \BibitemOpen
  \bibfield  {author} {\bibinfo {author} {\bibfnamefont {P.~E.}\ \bibnamefont
  {Bl\"ochl}},\ }\href {\doibase 10.1103/PhysRevB.50.17953} {\bibfield
  {journal} {\bibinfo  {journal} {Phys. Rev. B}\ }\textbf {\bibinfo {volume}
  {50}},\ \bibinfo {pages} {17953} (\bibinfo {year} {1994})}\BibitemShut
  {NoStop}%
\bibitem [{\citenamefont {Kresse}\ and\ \citenamefont
  {Hafner}(1993)}]{PhysRevB.47.558}%
  \BibitemOpen
  \bibfield  {author} {\bibinfo {author} {\bibfnamefont {G.}~\bibnamefont
  {Kresse}}\ and\ \bibinfo {author} {\bibfnamefont {J.}~\bibnamefont
  {Hafner}},\ }\href {\doibase 10.1103/PhysRevB.47.558} {\bibfield  {journal}
  {\bibinfo  {journal} {Phys. Rev. B}\ }\textbf {\bibinfo {volume} {47}},\
  \bibinfo {pages} {558} (\bibinfo {year} {1993})}\BibitemShut {NoStop}%
\bibitem [{\citenamefont {Kresse}\ and\ \citenamefont
  {Furthm\"uller}(1996)}]{PhysRevB.54.11169}%
  \BibitemOpen
  \bibfield  {author} {\bibinfo {author} {\bibfnamefont {G.}~\bibnamefont
  {Kresse}}\ and\ \bibinfo {author} {\bibfnamefont {J.}~\bibnamefont
  {Furthm\"uller}},\ }\href {\doibase 10.1103/PhysRevB.54.11169} {\bibfield
  {journal} {\bibinfo  {journal} {Phys. Rev. B}\ }\textbf {\bibinfo {volume}
  {54}},\ \bibinfo {pages} {11169} (\bibinfo {year} {1996})}\BibitemShut
  {NoStop}%
\bibitem [{\citenamefont {Dudarev}\ \emph {et~al.}(1998)\citenamefont
  {Dudarev}, \citenamefont {Botton}, \citenamefont {Savrasov}, \citenamefont
  {Humphreys},\ and\ \citenamefont {Sutton}}]{PhysRevB.57.1505}%
  \BibitemOpen
  \bibfield  {author} {\bibinfo {author} {\bibfnamefont {S.~L.}\ \bibnamefont
  {Dudarev}}, \bibinfo {author} {\bibfnamefont {G.~A.}\ \bibnamefont {Botton}},
  \bibinfo {author} {\bibfnamefont {S.~Y.}\ \bibnamefont {Savrasov}}, \bibinfo
  {author} {\bibfnamefont {C.~J.}\ \bibnamefont {Humphreys}}, \ and\ \bibinfo
  {author} {\bibfnamefont {A.~P.}\ \bibnamefont {Sutton}},\ }\href {\doibase
  10.1103/PhysRevB.57.1505} {\bibfield  {journal} {\bibinfo  {journal} {Phys.
  Rev. B}\ }\textbf {\bibinfo {volume} {57}},\ \bibinfo {pages} {1505}
  (\bibinfo {year} {1998})}\BibitemShut {NoStop}%
\bibitem [{\citenamefont {Dudarev}\ \emph {et~al.}(2018)\citenamefont
  {Dudarev}, \citenamefont {Liu}, \citenamefont {Andersson}, \citenamefont
  {Stanek}, \citenamefont {Ozaki},\ and\ \citenamefont
  {Franchini}}]{Dudarev2018}%
  \BibitemOpen
  \bibfield  {author} {\bibinfo {author} {\bibfnamefont {A.}~\bibnamefont
  {Dudarev}}, \bibinfo {author} {\bibfnamefont {P.}~\bibnamefont {Liu}},
  \bibinfo {author} {\bibfnamefont {D.}~\bibnamefont {Andersson}}, \bibinfo
  {author} {\bibfnamefont {C.}~\bibnamefont {Stanek}}, \bibinfo {author}
  {\bibfnamefont {T.}~\bibnamefont {Ozaki}}, \ and\ \bibinfo {author}
  {\bibfnamefont {C.}~\bibnamefont {Franchini}},\ }\href@noop {} {\bibfield
  {journal} {\bibinfo  {journal} {arXiv:}\ } (\bibinfo {year}
  {2018})}\BibitemShut {NoStop}%
\bibitem [{\citenamefont {Boykin}\ \emph {et~al.}(2007)\citenamefont {Boykin},
  \citenamefont {Kharche}, \citenamefont {Klimeck},\ and\ \citenamefont
  {Korkusinski}}]{Boykin}%
  \BibitemOpen
  \bibfield  {author} {\bibinfo {author} {\bibfnamefont {T.~B.}\ \bibnamefont
  {Boykin}}, \bibinfo {author} {\bibfnamefont {N.}~\bibnamefont {Kharche}},
  \bibinfo {author} {\bibfnamefont {G.}~\bibnamefont {Klimeck}}, \ and\
  \bibinfo {author} {\bibfnamefont {M.}~\bibnamefont {Korkusinski}},\ }\href
  {http://stacks.iop.org/0953-8984/19/i=3/a=036203} {\bibfield  {journal}
  {\bibinfo  {journal} {Journal of Physics: Condensed Matter}\ }\textbf
  {\bibinfo {volume} {19}},\ \bibinfo {pages} {036203} (\bibinfo {year}
  {2007})}\BibitemShut {NoStop}%
\bibitem [{\citenamefont {Popescu}\ and\ \citenamefont
  {Zunger}(2010)}]{Popescu2010}%
  \BibitemOpen
  \bibfield  {author} {\bibinfo {author} {\bibfnamefont {V.}~\bibnamefont
  {Popescu}}\ and\ \bibinfo {author} {\bibfnamefont {A.}~\bibnamefont
  {Zunger}},\ }\href@noop {} {\bibfield  {journal} {\bibinfo  {journal} {Phys.
  Rev. Lett.}\ }\textbf {\bibinfo {volume} {104}},\ \bibinfo {pages} {236403}
  (\bibinfo {year} {2010})}\BibitemShut {NoStop}%
\bibitem [{\citenamefont {Popescu}\ and\ \citenamefont
  {Zunger}(2012)}]{Popescu2012}%
  \BibitemOpen
  \bibfield  {author} {\bibinfo {author} {\bibfnamefont {V.}~\bibnamefont
  {Popescu}}\ and\ \bibinfo {author} {\bibfnamefont {A.}~\bibnamefont
  {Zunger}},\ }\href@noop {} {\bibfield  {journal} {\bibinfo  {journal} {Phys.
  Rev. B}\ }\textbf {\bibinfo {volume} {85}},\ \bibinfo {pages} {085201}
  (\bibinfo {year} {2012})}\BibitemShut {NoStop}%
\bibitem [{\citenamefont {Eckhardt}\ \emph {et~al.}(2014)\citenamefont
  {Eckhardt}, \citenamefont {Hummer},\ and\ \citenamefont
  {Kresse}}]{Eckhardt2014}%
  \BibitemOpen
  \bibfield  {author} {\bibinfo {author} {\bibfnamefont {C.}~\bibnamefont
  {Eckhardt}}, \bibinfo {author} {\bibfnamefont {K.}~\bibnamefont {Hummer}}, \
  and\ \bibinfo {author} {\bibfnamefont {G.}~\bibnamefont {Kresse}},\
  }\href@noop {} {\bibfield  {journal} {\bibinfo  {journal} {Phys. Rev. B}\
  }\textbf {\bibinfo {volume} {89}},\ \bibinfo {pages} {165201} (\bibinfo
  {year} {2014})}\BibitemShut {NoStop}%
\bibitem [{\citenamefont {Reticcioli}\ \emph {et~al.}(2016)\citenamefont
  {Reticcioli}, \citenamefont {Profeta}, \citenamefont {Franchini},\ and\
  \citenamefont {Continenza}}]{Reticcioli2015}%
  \BibitemOpen
  \bibfield  {author} {\bibinfo {author} {\bibfnamefont {M.}~\bibnamefont
  {Reticcioli}}, \bibinfo {author} {\bibfnamefont {G.}~\bibnamefont {Profeta}},
  \bibinfo {author} {\bibfnamefont {C.}~\bibnamefont {Franchini}}, \ and\
  \bibinfo {author} {\bibfnamefont {A.}~\bibnamefont {Continenza}},\
  }\href@noop {} {\bibfield  {journal} {\bibinfo  {journal} {Journal of
  Physics: Conference Series}\ }\textbf {\bibinfo {volume} {689}},\ \bibinfo
  {pages} {012027} (\bibinfo {year} {2016})}\BibitemShut {NoStop}%
\bibitem [{\citenamefont {Reticcioli}\ \emph {et~al.}(2017)\citenamefont
  {Reticcioli}, \citenamefont {Profeta}, \citenamefont {Franchini},\ and\
  \citenamefont {Continenza}}]{Reticcioli2017}%
  \BibitemOpen
  \bibfield  {author} {\bibinfo {author} {\bibfnamefont {M.}~\bibnamefont
  {Reticcioli}}, \bibinfo {author} {\bibfnamefont {G.}~\bibnamefont {Profeta}},
  \bibinfo {author} {\bibfnamefont {C.}~\bibnamefont {Franchini}}, \ and\
  \bibinfo {author} {\bibfnamefont {A.}~\bibnamefont {Continenza}},\ }\href
  {\doibase 10.1103/PhysRevB.95.214510} {\bibfield  {journal} {\bibinfo
  {journal} {Phys. Rev. B}\ }\textbf {\bibinfo {volume} {95}},\ \bibinfo
  {pages} {214510} (\bibinfo {year} {2017})}\BibitemShut {NoStop}%
\bibitem [{\citenamefont {Calder}\ \emph {et~al.}(2015)\citenamefont {Calder},
  \citenamefont {Lee}, \citenamefont {Stone}, \citenamefont {Lumsden},
  \citenamefont {Lang}, \citenamefont {Feygenson}, \citenamefont {Zhao},
  \citenamefont {Yan}, \citenamefont {Shi}, \citenamefont {Sun}, \citenamefont
  {Tsujimoto}, \citenamefont {Yamaura},\ and\ \citenamefont
  {Christianson}}]{Calder2015}%
  \BibitemOpen
  \bibfield  {author} {\bibinfo {author} {\bibfnamefont {S.}~\bibnamefont
  {Calder}}, \bibinfo {author} {\bibfnamefont {J.~H.}\ \bibnamefont {Lee}},
  \bibinfo {author} {\bibfnamefont {M.~B.}\ \bibnamefont {Stone}}, \bibinfo
  {author} {\bibfnamefont {M.~D.}\ \bibnamefont {Lumsden}}, \bibinfo {author}
  {\bibfnamefont {J.~C.}\ \bibnamefont {Lang}}, \bibinfo {author}
  {\bibfnamefont {M.}~\bibnamefont {Feygenson}}, \bibinfo {author}
  {\bibfnamefont {Z.}~\bibnamefont {Zhao}}, \bibinfo {author} {\bibfnamefont
  {J.-Q.}\ \bibnamefont {Yan}}, \bibinfo {author} {\bibfnamefont {Y.~G.}\
  \bibnamefont {Shi}}, \bibinfo {author} {\bibfnamefont {Y.~S.}\ \bibnamefont
  {Sun}}, \bibinfo {author} {\bibfnamefont {Y.}~\bibnamefont {Tsujimoto}},
  \bibinfo {author} {\bibfnamefont {K.}~\bibnamefont {Yamaura}}, \ and\
  \bibinfo {author} {\bibfnamefont {A.~D.}\ \bibnamefont {Christianson}},\
  }\href {https://doi.org/10.1038/ncomms9916} {\bibfield  {journal} {\bibinfo
  {journal} {Nature Communications}\ }\textbf {\bibinfo {volume} {6}},\
  \bibinfo {pages} {8916 EP } (\bibinfo {year} {2015})},\ \bibinfo {note}
  {article}\BibitemShut {NoStop}%
\bibitem [{\citenamefont {Mohn}(2006)}]{Mohn}%
  \BibitemOpen
  \bibfield  {author} {\bibinfo {author} {\bibfnamefont {P.}~\bibnamefont
  {Mohn}},\ }\href@noop {} {\emph {\bibinfo {title} {Magnetism in the Solid
  State}}}\ (\bibinfo  {publisher} {Springer, Berlin},\ \bibinfo {year}
  {2006})\BibitemShut {NoStop}%
\end{thebibliography}%

\end{document}